\documentclass[10pt,twocolumn,preprintnumbers]{article}
\usepackage{amssymb}
\usepackage{latexsym}
\usepackage{graphicx}
\usepackage{epstopdf}
\usepackage{caption}
\usepackage{graphicx}
\usepackage{bm}
\usepackage[latin1]{inputenc}
\usepackage{amsmath}
\usepackage{amsfonts}
\usepackage{subfigure}
\usepackage{amssymb}
\usepackage{makeidx}
\usepackage{color}
\usepackage{mathrsfs}
\RequirePackage[colorlinks,citecolor=blue,urlcolor=blue,linkcolor=magenta]{hyperref}
\newcommand{\jd}[1]{{\bf\textcolor{blue}{[JD:#1]}}}
\newcommand{\wk}[1]{{\bf\textcolor{magenta}{[WK:#1]}}}
\newcommand {\drm} {\dot{\rho}_{\rm mat}}
\newcommand {\rhm} {\rho_{\rm mat}}
\newcommand {\prm} {p_{\rm mat}}
\newcommand {\wm} {\omega_{\rm mat}}
\newcommand {\gtil} {\tilde{\gamma}}
\begin{document}
	
	\title{Linear growth index of matter perturbations in Rastall gravity}

	\author{
		Wompherdeiki Khyllep\thanks{sjwomkhyllep@gmail.com}${\ }^{a,b}$,
		Jibitesh Dutta\thanks{jibitesh@nehu.ac.in}${\ }^{c,d}$
		\\ 
		\normalsize{
			${}^a$ Department of Mathematics, North-Eastern Hill University, Shillong, Meghalaya 793022, India} \\
		\normalsize{
			${}^b$ Department of Mathematics, St. Anthony's College, Shillong, Meghalaya 793001, India }\\
		\normalsize{
			${}^c$ Mathematics Division, Department of Basic Sciences and Social Sciences, North-Eastern}\\ 
		\normalsize{Hill University,  Shillong, Meghalaya 793022, India} \\
		${}^d$ \normalsize{Inter University Centre for Astronomy and Astrophysics, Pune 411 007, India }\\
	}
	
	\date{}
	
	\maketitle

	\begin{abstract}
		Rastall gravity theory shows notable features consistent with physical observations in comparison to the standard Einstein theory. Recently, there has been a debate about the equivalence of  Rastall gravity and general relativity. Motivated by this open issue,  in the present work, we attempt to shed some light on this debate by analyzing the evolution of the Rastall based cosmological model at the background as well as perturbation level. Employing the dynamical system techniques, we found that at late times, the dynamics of the model resembles the $\Lambda$CDM model at the background level irrespective of the choice of  Rastall's parameter. However, at the perturbation level, we found that the evolution of the growth index heavily depends on the Rastall's parameter and displays a significant deviation from the $\Lambda$CDM model.
		
	\end{abstract}

	\section{Introduction}
	The inability of the standard theory of gravity to describe various observational evidence has led to the need for extending the theory of general relativity (GR). The structure of extended theories of gravity provides a useful approach to alleviate the fundamental problems such as dark matter (DM) and dark energy (DE) associated with the standard model of GR \cite{clifton,will,Nojiri:2017ncd}.
	
	One of the extended theories of gravity was proposed by Rastall in 1972 \cite{ras,ras1}. In his theory, Rastall assumed the violation of the conservation of energy-momentum tensor in curved spacetime without dropping the Bianchi identities. As the usual particle creation phenomenon violates the energy-momentum conservation law, therefore, this
	theory can also be regarded as a classical formulation of a particle creation process \cite{park1,park2,park3,hawk}.  Rastall gravity is one of the most promising non-conservative modified theory of gravity supporting various cosmological and astrophysical observations. Recently, it got a lot of attention in the physics community \cite{batis1,batis2,batis3,bro,dara}. For instance, this theory fits well with observational data related to the age of the Universe, Hubble parameter and helium nucleosynthesis \cite{raw1, raw}. Further, this theory can possibly circumvent the entropy and age problems of standard cosmology \cite{ker}. From Mach's principle perspective, this theory is more `Machian' than the standard theory of GR \cite{maj}.
	
	It has been claimed by Visser in  \cite{vis} that the theory of Rastall gravity is completely equivalent to GR and further supported by \cite{das} from a thermodynamical perspective. However, this result has been recently denied by \cite{dar,han} leaving an open debate that whether this theory is a modified gravity theory or it is equivalent to GR with an additional modified matter content. Earlier, the non-equivalence of Rastall gravity from GR has also been pointed by Smalley in Ref. \cite{sma}.
	
	In Ref. \cite{batis2} it has been shown that Rastall cosmology is equivalent to the $\Lambda$CDM at both background and linear perturbation level except that DE cluster for the former model.
	The growth of matter perturbations provides an efficient approach to predict the matter distribution of the Universe and also to discriminate various gravitational theories \cite{oka,ishak}. One of the simple observational tools used to study the growth history of a model is the so-called growth index of matter perturbations (denoted by $\gamma$) \cite{peb}.  The accurate estimation of the growth index is one of the basic tasks from the cosmological point of view as it can be used as a tool to test the validity of GR on extragalactic scales. It is a usual procedure that for each cosmological model, one requires to analyze its background evolution and the growth index of matter perturbations. This task helps one to get an overall effect of the model at the cosmological and astrophysical level. For instance, those DE models based on GR, the value of the constant growth index can be reduced to that of the $\Lambda$CDM model (i.e. $\gamma=\frac{6}{11}$) irrespective of the choice of model parameters. However, in the case of modified gravity based models, the value shows a significant deviation from $\frac{6}{11}$. Recently, there are various work in the literature where an analytical form of the growth index for different models are obtained which include the scalar field DE \cite{waga,wang,ness}, DGP\cite{lue,gong}, $f(R)$ gravity \cite{gan,gan1,mena}, Finsler-Randers \cite{basi1}, time-varying vacuum models \cite{basi2}, clustered DE \cite{basi3}, holographic DE \cite{basi4} and $f(T)$ gravity \cite{basi5}.

	At the level of linear perturbation, it has been obtained that the evolution equation of the growth of matter perturbations for Rastall based cosmological model coincides with that of the $\Lambda$CDM model. Additionally, in the case of Rastall model, DE perturbations cluster even if it is in the vacuum energy form \cite{batis2}. This might have a strong impact on the value of the growth index, leading to interesting cosmological signatures of the model which can be tested with growth rate data. It is therefore interesting to analyze the growth index of the Rastall model in order to shed some light on the said debate.
	
	With this motivation the aim of the paper is two-fold. First, we analyze the dynamics of the Rastall model at the background level using the dynamical system techniques. The dynamical system analysis is performed in Sec.  \ref{sec:ds_rgc} to qualitatively extract the evolution described by the basic cosmological equations presented in Sec. \ref{sec:bce_rgc}. Secondly, we investigate the behavior of this model at the linear perturbation level in Sec. \ref{sec:linear_growth}. For this we shall derive the growth index of the Rastall model and then compare with that of  the $\Lambda$CDM in subsections \ref{subsec:cons_g_ind} and \ref{subsec:vary_g_ind}. Finally, the conclusion is given in Sec. \ref{sec:conc}.

	\section{Basic cosmological equations of Rastall gravity}\label{sec:bce_rgc}
	In this section, we briefly introduce the basic cosmological equations of Rastall theory of gravity.  In many gravitational theories, the source of energy-momentum is determined by a vanishing divergence tensor, minimally coupled to the geometry. However, in this theory,  the usual conservation law of the energy-momentum tensor is not satisfied, instead, it is assumed to satisfy the following relation \cite{ras,ras1}
	{\small\begin{eqnarray}\label{non_conservation}
		T^{\mu \nu}_{;\mu}&=&\frac{\lambda}{8\pi G} R^{;\nu}\, ,
		\end{eqnarray}}
	
	\noindent where semi-colon denotes the covariant derivative, $G$  is the Newton's gravitational constant, $T_{\mu\nu}$ is the energy-momentum tensor and $R$ is the Ricci scalar. Here,  $\lambda$ is the coupling constant which measure the exchange of energy between geometry and matter field. When $\lambda=0$, one obtain a usual conservation law in GR. The modified Einstein's field equations in Rastall gravity framework are given by
	{\small \begin{eqnarray}
		G_{\mu \nu}&=&8 \pi G \Big(T_{\mu\nu}-\frac{\gtil-1}{2} g_{\mu\nu} T \Big)\, , \label{EFE_1_1}\\
		T^{\mu\nu}_{;\mu}&=&\frac{\gtil-1}{2}T^{;\nu}\, , \label{EFE_1_2}
		\end{eqnarray}}
	
	\noindent where $T$ is the trace of the energy-momentum tensor and $\gtil$ is the Rastall's parameter which is related to $\lambda$ as
	{\small \begin{eqnarray}
		\gtil&=&\frac{1+6\lambda}{1+4\lambda}\,.
		\end{eqnarray}}
	Clearly, $\gtil=1$ corresponds to the GR case.
	
	We now consider a two perfect fluid system consisting of a pressureless matter with corresponding energy density $\rho_m$ and an exotic component with energy density $\rho_x$ which accounts for acceleration of the universe and whose equation of state (EoS) is  $w_x$.  It is worth mentioning here that an extra exotic fluid $\rho_x$ behaving as a source of DE is required similar to the GR case \cite{batis2}. Further, we assume that the matter component satisfies the usual conservation law. Therefore, the modified Einstein's field equations \eqref{EFE_1_1}-\eqref{EFE_1_2} become
	{\small \begin{eqnarray}
		G_{\mu \nu}&=&8 \pi G \Big(T^m_{\mu\nu}+T^x_{\mu\nu}-\frac{\gtil-1}{2} g_{\mu\nu} (T^m+T^x) \Big)\, , \label{EFE_2}\\
		T^{\mu\nu}_{;\mu}&=&\frac{\gtil-1}{2}(T_x^{;\nu}+T_m^{;\nu})\,, \,\,\, ~~~ T_{m;\mu}^{\mu\nu}=0\,,\label{EFE_3}
		\end{eqnarray}}
	
	\noindent where index $m$ denotes the matter component and $x$ denotes the DE component. 
	
	At large scale, various observational data favor a homogeneous and isotropic universe which can be described by a spatially flat Friedmann-Lemaitre-Robertson-Walker (FLRW) metric given by
	{\small \begin{equation}\label{metric}
		ds^2=g_{\mu\nu}dx^\mu dx^\nu\,,~~~~(\mu,\nu=0,1,2,3)
		\end{equation}}
	
	\noindent with $g_{\mu\nu}=$diag $(1,-a^2(t),-a^2(t)\,r^2,-a^2(t)\,r^2\,\sin^2 \theta)$. Here, $a(t)$ is a scale factor and $t$ denotes the cosmological time. The above Einstein field equations \eqref{EFE_2}, \eqref{EFE_3} with respect to the metric \eqref{metric} can be written as
	{\small \begin{eqnarray}
		H^2& =&  \frac{4 \pi G}{3} \Big[\rho_x\,(3-\gtil-3(1-\gtil) w_x)  \nonumber\\&& +(3-\gtil) \rho_m\Big]\, , \label{frd_eqn}\\
		\dot{H}+H^2&  = & \frac{4 \pi G}{3} \Big[\rho_x\,(3(\gtil-2) w_x-\gtil)-\gtil \rho_m\Big],~~~~ \label{ryc_eqn}\\
		\dot{\rho}_x+3H (1+w_x) \rho_x&=&\frac{\gtil-1}{2} \Big[(1-3w_x) \dot{\rho}_x+\dot{\rho}_m\Big]\,, \label{de_cons}\\
		\dot{\rho}_m+3H \rho_m&=&0\,, \label{dm_cons}
		\end{eqnarray}}
	
	\noindent where $H=\frac{\dot{a}}{a}$ is the Hubble parameter and the overdot denotes derivative with respect to $t$. On solving Eqs. \eqref{de_cons} and \eqref{dm_cons}, one can obtain the evolution equation of $\rho_m$ and $\rho_x$ as \cite{bro}
	{\small \begin{eqnarray}
		\rho_x&=&\rho_{d 0}\, a^{-\frac{6(1+w_x)}{2+(1-\gtil)(1-3w_x)}}+\frac{(1-\gtil)\rho_{m0} a^{-3}}{2w_x+(\gtil-1)(1-3w_x)},~~~~ \label{de_dens}\\
		\rho_m&=&\rho_{m0} a^{-3}, \label{dm_dens}
		\end{eqnarray}}
	where $\rho_{d 0}$ is given by 
	{\small \begin{eqnarray}
		\rho_{d0}&=&\rho_{x0}+\frac{\gtil-1}{2w_x+(\gtil-1)(1-3w_x)} \rho_{m0}. \label{pres_dens}
		\end{eqnarray}}
	
	\noindent Eq. \eqref{de_dens} shows that the energy density of an exotic fluid consists of two components and one of them behaves as DM. In the above equations $\rho_{m0}$, $\rho_{x0}$ and $\rho_{d0}$ denote the present value of matter density, of an exotic fluid  and of a DE component of an exotic fluid.  With the aid of \eqref{de_dens}, \eqref{dm_dens} and \eqref{pres_dens}, the Friedmann equation \eqref{frd_eqn} can be written as
	{\small \begin{eqnarray}
		E^2(a)& \textstyle = & \textstyle \Big[\Big( \frac{A}{2} \Omega_{m0} +\frac{B}{2}\Omega_{x0}  \Big)\, \textstyle  {a}^{-{\frac {6(1+w_x)}{B}}}+\frac {A}{2}\Omega_{m0} {a}^{-3} \Big] \nonumber\\&&   \textstyle +\,\frac { \left( 3-\gtil
			\right) }{2}\,\Omega_{m0}\,{a}^{-3}\,,  \label{Ea}\\
		\text{where}~  A&=&\frac{B (\gtil-1)}{2(w_x+1)-B}, ~B=2-(\gtil-1)(1-3w_x).\nonumber
		\end{eqnarray}}
	
	\noindent In the above equation, $E(a)=\frac{H(a)}{H_0}$,  $\Omega_{m0}=\frac{8 \pi G \rho_{m0}}{3H_0^2}$, $\Omega_{x0}=\frac{8 \pi G \rho_{x0}}{3H_0^2}$ and $H_0$ is the present value of the Hubble parameter. Note that the expression within the square bracket of Eq. \eqref{Ea} denotes the energy contribution from an exotic fluid with one of its components behaving as DM. The term outside the square bracket denotes the contribution solely from a pressureless matter fluid.  As expected,  for $\gtil=1$,  the above equation reduces to the corresponding GR case given by
	{\small \begin{eqnarray}
		E_{GR}^2(a)& \textstyle = & \textstyle \Omega_{x0}  \, \textstyle  {a}^{-3(1+w_x)}+\Omega_{m0}\,{a}^{-3}\,, 
		\end{eqnarray}}
	with  $x$-component playing the role of DE  and $m$-component plays the role of DM. On differentiating \eqref{Ea} with respect to the scale factor, we obtain
	
	{\small \begin{equation}\label{d_lnE_d_lna}
		\frac{d \ln E}{d \ln a}=-\frac{3 \, \Omega_m\,w_x\,(2\gtil-3)}{3\gtil\,w_x-\gtil-3w_x+3}\,.
		\end{equation}}
	
	Further, from Eqs. \eqref{frd_eqn} and \eqref{ryc_eqn}, we can define the effective energy density and pressure as
	{\small \begin{eqnarray}
		p_{\rm eff}&=&\frac{\rho_x}{2}(\gtil-1-w_x(3\gtil-5))+\frac{\gtil-1}{2}\rho_m\,, \\
		\rho_{\rm eff}&=&\frac{\rho_x}{2} (3-\gtil-3(1-\gtil) w_x)+\frac{3-\gtil}{2}\rho_m\,.
		\end{eqnarray}}
	Then the effective EoS is given by
	{\small \begin{align}\label{w_eff}
		w_{\rm eff}=\frac{p_{\rm eff}}{\rho_{\rm eff}}=-1+\frac{2}{3-\gtil} [1+(3-2\gtil)w_x \, \Omega_x ].
		\end{align}}

	For accelerated universe, we must have $w_{\rm eff}<-\frac{1}{3}$.  In order to obtain a concrete picture on the background cosmological evolution described by this model,  in the next section, we shall perform a dynamical system analysis for this model.

	\section{Dynamical system analysis of Rastall model} \label{sec:ds_rgc}
	In this section, we shall perform a complete dynamical system analysis of the Rastall gravity based model,  where we introduce suitable dimensionless variables to recast the cosmological equations \eqref{frd_eqn}-\eqref{dm_cons} into an autonomous dynamical system. In this case, we choose the dimensionless energy density parameters of matter and of an exotic fluid given by
	{\small \begin{align}
		\Omega_m=\frac{8 \pi G \rho_m}{3H^2},~~~ \Omega_x=\frac{8 \pi G \rho_x}{3H^2}\,,
		\end{align}}
	
	\noindent respectively, as the dynamical variables of the system which are related by Eq. \eqref{frd_eqn} as
	{\small \begin{align}\label{constraint}
		\Omega_x (3-\gtil-3(1-\gtil)w_x)+(3-\gtil) \Omega_m=2.
		\end{align}}
	  It may be noted that depending on the choice of parameters $\gtil$ and $w_x$, there is a possibility that either $\Omega_m$ or $\Omega_x$ or both assume negative values. This feature usually arises in cosmological models where a phenomenological  interaction between different components is present and it is also related to an uncertainty in defining the relative energy densities of the model \cite{Quartin:2008px}.  Since in this Rastall cosmological model the exotic fluid $\rho_x$ and matter fluid $\rho_m$ can be interpreted as effectively interacting quantities (cf. Eq. \eqref{de_cons}), it is therefore not surprising that  $\Omega_m$ or $\Omega_x$ take negative values. Further, it is important to note here that the condition $\Omega_x<0$ is actually obligated in several $f(R)$ theories in order to alleviate the coincidence problem and also to match with the current observations \cite{jdbar}. While one could argue that $\Omega_m<0$ is physically unacceptable, the present model demands this condition (cf. Eq. \eqref{constraint}) and discarding it may lead to incomplete dynamics. For relevant recent work one can also see \cite{zonun,Dutta:2017fjw}. Again, it is  interesting to observe that negative energy density does not always imply phantom behavior in general (one can see \cite{Quartin:2008px}, \cite{Dutta:2017fjw}). For instance, in the present model, even if negative energy density is allowed, it is possible that both the DE component ($x$-component) and the overall behavior is non-phantom i.e. $w_x>-1$ and $w_{\rm eff}>-1$. This can be seen from Eq. \eqref{w_eff} (for $\tilde{\gamma}=\frac{3}{2}$,  $w_{\rm eff}>-1$ even if $\Omega_x<0$). As expected for $\gtil=1$, the relation \eqref{constraint} reduces to that of the GR case i.e. $\Omega_m+\Omega_x=1$.

	\begin{figure}[h!]
		\centering
		\includegraphics[width=6cm,height=3.5cm]{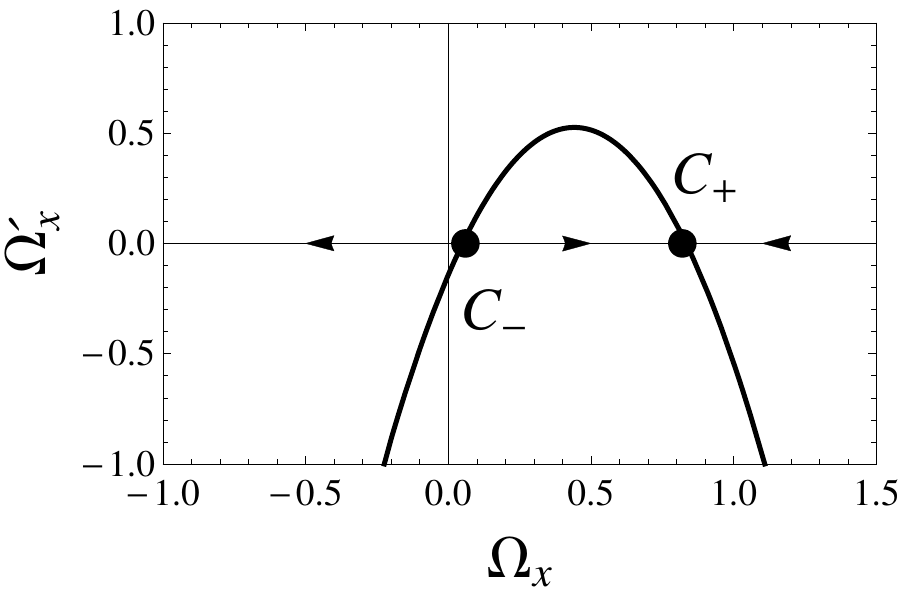}
		\caption{The phase portrait of the system \eqref{ode} describing the stability of critical points of the system. Here, $\gtil=0.9$ and $w_x=-1$.}
		\label{fig_ph_port_bg}
	\end{figure}
	
	\begin{figure}[h!]
		\centering
		\includegraphics[width=6cm,height=3.5cm]{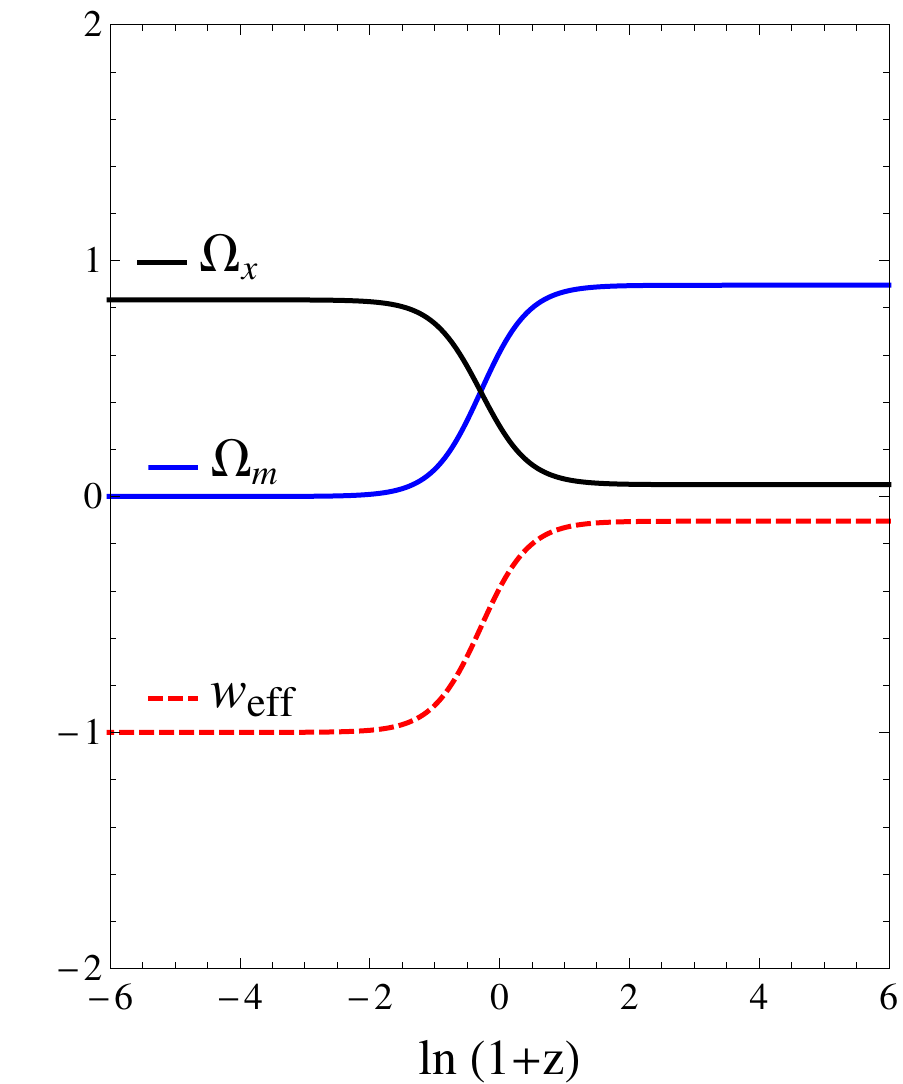}   \label{fig_bcg_para_quin}
		\caption{The evolution of the relative matter energy density $\Omega_{m}$, the relative DE density $\Omega_x$ and effective EoS $w_{\rm eff}$. Here,  $\gtil=0.9$ and $w_x=-1$.}
		\label{fig_bcg_para}
	\end{figure}

	From the above constraint equation \eqref{constraint}, it can be seen that one of the dimensionless density parameters  is linearly independent. Then the above governing cosmological equations \eqref{frd_eqn}-\eqref{dm_cons} can be recast into a following one-dimensional system:
	{\small \begin{eqnarray}\label{ode}
		\Omega_x'&=&\frac{3}{2 (\gtil-3)}\, \Big(3 w_x \Omega_x \left( \gtil^2-4\Omega_x+3\right)+8 \gtil w_x\,\Omega_x \nonumber\\&&\times \left(\Omega_x-1\right)-\Omega_x (\gtil^2+1)+2\gtil (\Omega_x-1)+2\Big)\,,
		\end{eqnarray}}
	
	\noindent where prime denotes a derivative with respect to a new time variable $N= \ln a$. The critical points for this one-dimensional system are
	$\Omega_x=\frac{1}{8\,{w_x}\, \left( 2\,\gtil-3 \right) }\,\Big[\Delta\pm\Big(\Delta^2-32\,w_x (\gtil-1)\,(2\gtil-3)\Big)^\frac{1}{2}\Big]$ where $\Delta=-{w_x}\, \left( 2\,\gtil-3 \right) ^{2}+ \left( \gtil-1 \right) ^
	{2}+\gtil\,{w_x}\, \left( \gtil-4 \right)$. We denote these critical points by $C_{\pm}$. On linearizing Eq. \eqref{ode} around critical points $C_+$ and $C_-$, one obtains small perturbations $\delta_+$, $\delta_-$ respectively as
	{\small \begin{eqnarray}\label{bg_pert}
		\delta_+'&=&\lambda_+\,\delta_+ \Rightarrow \delta_+(N)=\bar{\delta}_+ \exp^{\lambda_+ N}\\
		\delta_-'&=&\lambda_-\,\delta_- \Rightarrow \delta_-(N)=\bar{\delta}_- \exp^{\lambda_- N}
		\end{eqnarray}}
	
	\noindent where  $\lambda_\pm=\pm\,\frac{3}{2\,(\gtil-3)}\,\Big(\Delta^2-32\,w_x (\gtil-1)\,(2\gtil-3)\Big)^\frac{1}{2}$. Here, $\bar{\delta}_+$, $ \bar{\delta}_-$ denote the values of ${\delta}_+$, ${\delta}_-$ when $N=0$ respectively.  Therefore, for $\gtil>3$ case, the perturbation corresponds to critical point $C_+$ increases exponentially, but for $C_-$ the corresponding perturbation decreases exponentially. The situation is vice versa for $\gtil<3$ case. Hence, point $C_+$ (point $C_-$) corresponds to a stable (unstable) node for    $\gtil<3$ and unstable (stable) node for  $\gtil>3$.  Therefore, the local stability of critical points of the system undergoes an abrupt change as we vary $\gtil$ from a value smaller than 3 to greater than 3. In particular, the two critical points exchange their stability behavior as $\gtil$ passes through a value $\gtil=3$.  Hence, the system seems to undergo a bifurcation similar to the so-called {\it transcritical} bifurcation except that in the present scenario the critical points of the system vanish at $\gtil=3$. This occurs due to the fact that the system \eqref{ode} blows up at $\gtil=3$ and therefore, we cannot extract any dynamics at the background level using the present dynamical variables. Interestingly, for $\gtil=\frac{3}{2}$, we have checked that the above system contains only one unstable critical point which corresponds to a radiation dominated universe ($w_{\rm eff}=\frac{1}{3}$). However, from Eq. \eqref{de_dens} the case where $w_x=-1$ and $\gtil=\frac{3}{2}$ is undesirable as it leads to divergence of $\rho_x$.  Further, it can be seen that for $\gtil \simeq 1$ and $w_x \simeq -1$, point $C_+$ corresponds to $\Omega_x \simeq 1$ and point $C_-$ corresponds to $\Omega_x \simeq 0$. In general, for $\gtil<3$ (except $\gtil=\frac{3}{2}$), point $C_+$ corresponds to a DE dominated universe and $C_-$ corresponds to a matter dominated universe (as $\Omega_x$ for $C_+$ dominates that of $C_-$ and $\Omega_m$ for $C_-$ dominates that of $C_+$). Furthermore, for $\gtil>3$, point $C_-$ corresponds to a DE dominated universe and $C_+$ corresponds to a matter dominated universe. Hence, irrespective of value of $\gtil$ this model describes the late time transition of the universe from an unaccelerated DM to an accelerated DE domination at the background level. We have indeed checked that there are no other critical points that are hidden towards an infinite regime of $\Omega_x$. A particular case of the phase portrait for the system \eqref{ode} is given in Fig. \ref{fig_ph_port_bg} for $\gtil=0.9$, $w_x=-1$. It is also interesting to note that for $w_x=-1$, the late time attractor corresponds to $w_{\rm eff} = -1$ for any choices of $\gtil$. Thus, the background evolution of this model coincides exactly with the $\Lambda$CDM model at late  times, however the evolution during the matter domination depends on the value of the parameter $\gtil$.  This behavior is depicted numerically  in Fig. \ref{fig_bcg_para} by plotting the evolution of the DE and DM energy density parameters along with an effective EoS against redshift $z$ (where $1+z=\frac{a_0}{a}$ and $a_0$ is the present value of scale factor taken to be 1).  In the next section, we shall analyze the behavior of this model at the linear perturbation level and compare the evolution with the $\Lambda$CDM model.

	\section{Analysis of growth index of matter perturbations} \label{sec:linear_growth}
	
	Here, we shall study the growth of linear matter fluctuations of the present model in the matter dominated epoch. We first present the evolution equation of the matter perturbations of first order within the sub-horizon scale, then we study the evolution of growth index of matter perturbations for this model. Following \cite{batis2,batis3}, we consider the perturbed metric with synchronous gauge  given by
	{\small \begin{eqnarray}
		\bar{g}_{\mu\nu}&=&g_{\mu\nu}+h_{\mu\nu},~~~h_{\mu 0}=0\,,  \label{pert_metr}
		\end{eqnarray} }
	
	\noindent where $\bar{g}_{\mu\nu}$ is the total metric  and $h_{\mu\nu}$ is a small fluctuation of a background metric  $g_{\mu\nu}$ given by Eq. \eqref{metric}. For analysis of growth index, we consider  the following physical quantities to be perturbed viz. the energy density and pressure of an $x$-fluid ($\rho_x$, $p_x$), the energy density of a matter ($\rho_m$), the four-velocity of an $x$-fluid and matter ($u_x$, $u_m$) as 
	{\small\begin{eqnarray}
		\bar{\rho}_x&=&\rho_x+\delta \rho_x,~~~\bar{\rho}_m=\rho_m+\delta \rho_m,~~~\bar{p}_x=p_x+\delta p_x,~~~ \nonumber\\ \bar{u}_x&=&u_x+\delta u_x,~~~\bar{u}_m=u_m+\delta u_m.
		\end{eqnarray}}
	
	\noindent The basic differential equation describing the evolution  of the matter overdensities $\delta_m$ defined as $\frac{\delta \rho_m}{\rho_{m}}$  is given by
	{\small\begin{eqnarray}
		\ddot{\delta}_m+2 \nu H \dot{\delta}_m-4 \pi G \mu \rho_m\delta_m=0\,. \label{pert_ov_dens}
		\end{eqnarray}}
	
	\noindent  In the above equation, the quantities $\nu$ and $\mu$ parametrizes the deviation of any gravitational theory from GR. In the case of DE models within the GR framework, we have $\mu=\nu=1$, however, in the case of modified gravity models, we have $\nu=1$ and $\mu \neq 1$.  On using the perturbed metric \eqref{pert_metr} in this model, we get
	{\small \begin{eqnarray}
		\nu&=&1~~ \text{and} ~~\mu=\gtil+\left(\gtil+3(2-\gtil)\,w_x\right) \frac{\delta_x}{\delta_m}\,\frac{\rho_x}{\rho_m}, \label{mu_nu}
		\end{eqnarray}}
	
	\noindent where  $\delta_x=\frac{\delta \rho_x}{\rho_{x}}$. For a detailed and complete perturbed equations of this model, one can see \cite{batis2,batis3}. In order to obtain the quantity $\mu$, we first need to determine the functional form of $\frac{\delta_x}{\delta_m}$. For this we shall assume an adiabatic initial condition of density perturbation given by \cite{bal,lyt,bau,amsu}
	{\small \begin{equation}\label{adia_ini}
		\frac{\delta \rho_x}{\dot{\rho}_x}=\frac{\delta \rho_m}{\dot{\rho}_m}.
		\end{equation}  }
	The above adiabatic condition is defined   without a prior choice  of the underlying gravitational theory \cite{lyt}, \cite{bau}. Further, we note that adiabatic condition is in agreement with observations and simple inflationary models usually predict adiabatic perturbations \cite{Netterfield:2001yq}. 
	Using Eqs. \eqref{de_cons}, \eqref{dm_cons}, the condition \eqref{adia_ini} can be written as
	{\small \begin{equation}\label{adia_ini_1}
		\frac{\delta_x}{\delta_m}=\frac{\frac{\gtil-1}{2}\frac{\Omega_m}{\Omega_x}+(1+w_x)}{1+\frac{1-\gtil}{2}(1-3w_x)}.
		\end{equation} }
	Interestingly,  for $\gtil=1$, we get the corresponding result of GR \cite{bal} i.e.
		\begin{align}
		\frac{\delta_x}{\delta_m}=1+w_x\,.
		\end{align}
		 On employing \eqref{adia_ini_1}, we see that the evolution equation of matter perturbations \eqref{pert_ov_dens} for $w_x=-1$, coincides with that of the $\Lambda$CDM model irrespective of the choice of $\gtil$. This is in agreement with the result obtained in \cite{batis2}. The only difference in this model in comparison to the $\Lambda$CDM  is that  an extra relation which  relates DE and DM perturbations arises which is given  by  \cite{batis2}
	{\small \begin{equation}\label{dm_de_pert_reln}
		\delta \rho_x=\frac{\gtil-1}{2(3-2\gtil)} \delta \rho_m.
		\end{equation} }
	
	\noindent Therefore, DE clusters even though it is in the vacuum energy form. This suggest that the parameter $\gtil$ might somehow affect the behavior of the growth index. It worth mentioning here that the condition \eqref{adia_ini_1} reduces to Eq. \eqref{dm_de_pert_reln} for $w_x=-1$ and which is indeed consistent with a full set of linear perturbation equations of this model presented in \cite{batis2}. Note that in Eq. \eqref{dm_de_pert_reln}, the case $\gtil= \frac{3}{2}$ is not feasible as discussed earlier.  On employing the favorable choice of parameters values $\gtil \simeq 1.41$, $w_x \simeq -1$ and $\Omega_{m}\simeq 0.296$  \cite{batis3} on Eq. \eqref{adia_ini_1},  we obtain $\frac{\delta_x}{\delta_m} \simeq 0.08$ which is in agreement with those predicted by previous studies for clustered DE models \cite{Basilakos:2014yda}. Now, we shall focus on the investigation of the growth index of matter perturbations denoted by $\gamma$. In this regard, we consider a  growth rate of clustering $f$ given by \cite{peb}
	{\small \begin{equation} \label{gr_ind}
		f(a)=\frac{d \ln \delta_m}{d \ln a}\simeq \Omega^\gamma_m (a)\,,
		\end{equation}}
	
	\noindent where
	
	{\small \begin{equation} \label{Om_Om0}
		\Omega_m(a)=\frac{\Omega_{m0}\,a^{-3}}{E^2(a)}.
		\end{equation}}
	
	\noindent The growth index $\gamma$ is used to distinguish a modified gravity theory from the standard GR theory  on cosmological scales. On differentiating Eq. \eqref{Om_Om0}, one can obtain the equation
	{\small \begin{equation}\label{d_Om_d_a}
		\frac{d \Omega_m}{d a}=-3 \frac{\Omega_m(a)}{a}\Big(1+\frac{2}{3}\,\frac{d \ln E}{d\ln a}\Big).
		\end{equation}}
	
	\noindent Using the first equality of \eqref{gr_ind}, the Eq. \eqref{pert_ov_dens} can be written as
	{\small \begin{equation}\label{df_da}
		a\frac{d f}{d a}+\left(2 \nu+\frac{d \ln E}{d \ln a} \right)\,f+f^2=\frac{3 \mu\,\Omega_m}{2}.
		\end{equation}}
	In what follows, we analyze the behavior of growth index $\gamma$ for two cases separately.
	
	\begin{figure}[h!]
		\centering
		\subfigure[]{%
			\includegraphics[width=6cm,height=3.5cm]{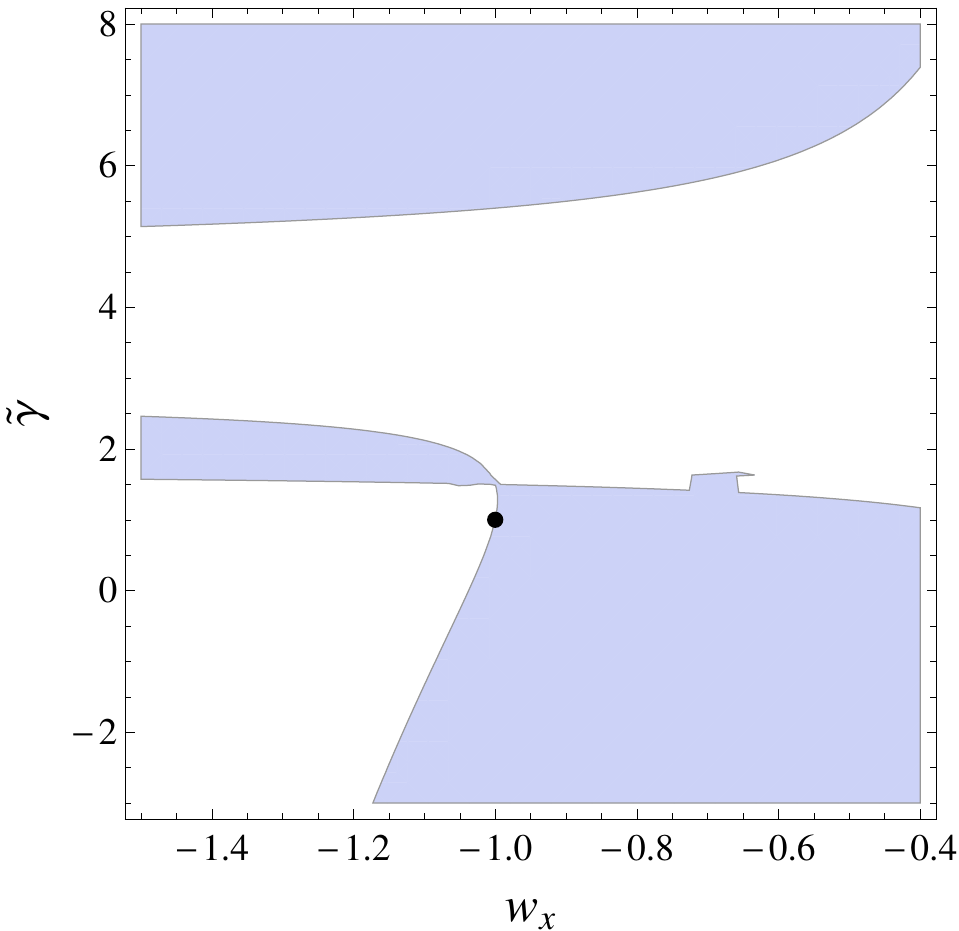}\label{fig_fix_gr_ind_reg_plot}}
		\qquad
		\subfigure[]{%
			\includegraphics[width=6cm,height=3.5cm]{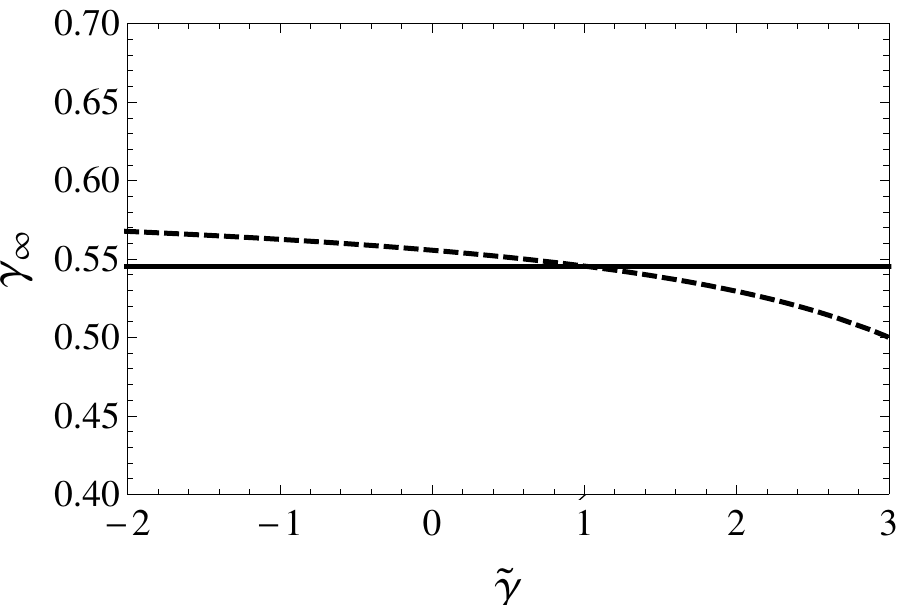}\label{fig_fix_gr_ind_ras_lcdm}}
		\qquad
		\subfigure[]{%
			\includegraphics[width=6cm,height=3.5cm]{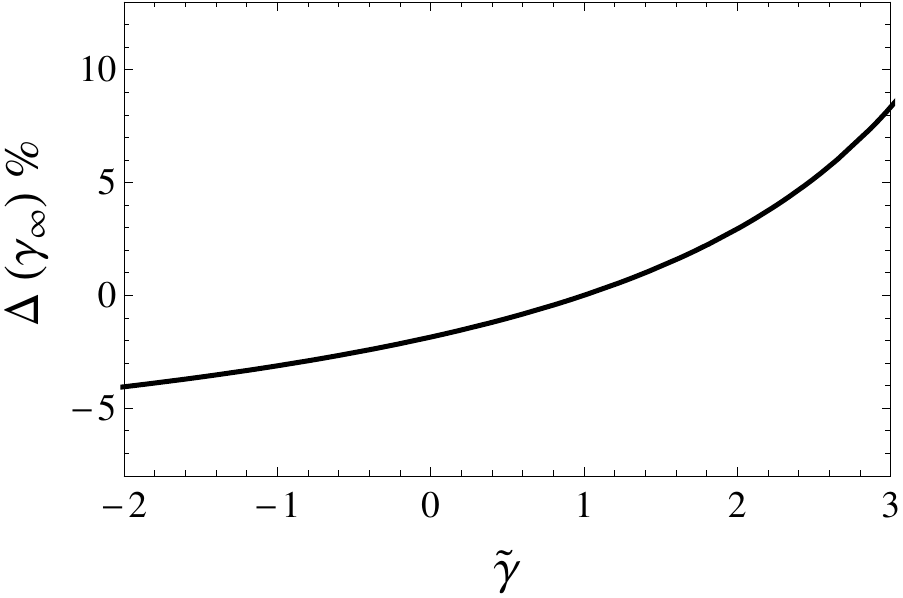}\label{fig_fix_gr_ind_ras_lcdm_percent}}
		\caption{(a) The region in $(w_x,\gtil)$ parameter space where the  asymptotic value of  growth index $\gamma_\infty$ is greater than that of the $\Lambda$CDM model. The solid circle corresponds to the $\Lambda$CDM model. (b) The evolution of $\gamma_\infty$ for Rastall's model  as a function of $\gtil$ (in dashed curve). The solid line corresponds to the $\Lambda$CDM model. (c) The relative difference $\Delta(\gamma_\infty)=[1-\gamma_\infty/\gamma_\infty^{(\Lambda)}] \%$ of the asymptotic value of growth index for the Rastall's model with respect to the $\Lambda$CDM model. In (b) and (c), we have taken $w_x=-1$.} 
		\label{fig_const_gr_ind_ras_lcdm}
	\end{figure} 
	
	\subsubsection{Constant growth index} \label{subsec:cons_g_ind}
	In this subsection, we consider the simplest choice of the growth index i.e. the asymptotic value of growth index denoted by $\gamma_\infty$. An analytical  approach for finding the asymptotic growth index was developed by Steigerwald {\it et al.}  \cite{stei}. An application of an asymptotic value of growth index to various class of DE models was discussed in  \cite{basi2}. Within an analytical method developed by Steigerwald {\it et al.}, the asymptotic value of growth index $\gamma_\infty$ is given by
	{\small \begin{equation}
		\gamma_\infty=\frac{3(M_0+M_1)-2(H_1+N_1)}{2+2X_1+3M_0}\,,
		\end{equation}}
	
	\noindent where the above quantities are defined as
	{\small \begin{eqnarray}
		M_0&=&\mu \vert_{\omega=\Theta},~~~M_1=\frac{d \mu}{d \omega}\Big\vert_{\omega=\Theta}, ~~~ N_1=\frac{d \nu}{d \omega}\Big\vert_{\omega=\Theta},  \nonumber\\[2ex]
		H_1&=&-\frac{X_1}{2}=\frac{d(\ln E/d \ln a)}{d \omega}\Big\vert_{\omega=\Theta}\,,
		\end{eqnarray}}
	
	\noindent with $\omega=\ln \Omega_m(a)$ and $\Theta=\ln\left(\frac{2}{3-\gtil}\right)$. It is worth mentioning that the growth index is determined within the matter dominated epoch which corresponds to $\Omega_x=0$. According to relation \eqref{constraint}, this corresponds to $\Omega_m=\frac{2}{3-\gtil}$. After some algebraic calculations, based, on Eqs. \eqref{d_lnE_d_lna}, \eqref{mu_nu}, \eqref{adia_ini_1} we get
	{\small \begin{eqnarray}
		M_0&=&\frac {2\,(3\,\gtil\,w_x+\gtil-3\,w_x)}{3\,\gtil\,w_x-\gtil-3\,w_x+3}\,,~~~ N_1=0\,,\nonumber \\
		M_1&=&\frac {-2(1+w_x)(\gtil - 3)(3\gtil w_x-\gtil - 6 w_x)}{ \left( 3\,
			\gtil\,w_x-\gtil-3\,w_x+3 \right) ^{2}}\,, \nonumber \\
		H_1&=&\frac {-6\,w_x \left( 2\,\gtil-3 \right) }{ \left( 3-\gtil \right) 
			\left( 3\,\gtil\,w_x-\gtil-3\,w_x+3 \right) }\,.
		\end{eqnarray}}
	
	\noindent Thus, the value of the asymptotic growth index $\gamma_\infty$ is given by
	\begin{equation}\label{gam_inf_rast}
	\textstyle \gamma_\infty=\frac {3 \,w_x \left( 2\,\gtil-3 \right)  \left( 3\,{\gtil}^{2}w_x-{
			\gtil}^{2}-12\,\gtil\,w_x+14\,\gtil-3\,w_x-33 \right) }{ \left( 12\,{
			\gtil}^{2}w_x+2\,{\gtil}^{2}-72\,\gtil\,w_x-3\,\gtil+72\,w_x-9 \right) 
		\left( 3\,\gtil\,w_x-\gtil-3\,w_x+3 \right) }\,.
	\end{equation}
	For $w_x=-1$, we get $\gamma_\infty=\frac{3(\gtil-5)}{5 \gtil-27}$, which interestingly depends on $\gtil$ even though the evolution equation for matter perturbations (i.e. Eq. \eqref{pert_ov_dens}) coincides with the corresponding equation for the standard $\Lambda$CDM model.  As expected, the above formula reduces to its standard value for $\Lambda$CDM model i.e $\gamma^{(\Lambda)}_\infty=\frac{6}{11}$ for $\gtil=1$ and $w_x=-1$. In Fig. \ref{fig_fix_gr_ind_reg_plot}, we plot the region  in the $(w_x,\gtil)$ parameter space where the  asymptotic value of  growth index $\gamma_\infty$ is greater than that of the $\Lambda$CDM model [i.e. $\gamma_\infty^{(\Lambda)}=\frac{6}{11}$]. In Fig. \ref{fig_fix_gr_ind_ras_lcdm}, we plot the asymptotic value of growth index as a function of a parameter $\gtil$ for $w_x=-1$. It can be concluded that when $w_x=-1$, the value of $\gamma_\infty$ for this model is greater than that of $\Lambda$CDM for $\gtil<1$ or $\gtil>\frac{27}{5}$.  The relative deviation of $\gamma_\infty$ for this model with respect to the $\Lambda$CDM model as a function of $\gtil$ is given in Fig. \ref{fig_fix_gr_ind_ras_lcdm_percent}.
	
	Further, by using the favorable choice of parameters values $\gtil=1.41$, $w_x=-1$ and $\Omega_{m0}=0.296$ \cite{batis3},  we obtain $\gamma_\infty=0.5304$. Therefore, we see that the asymptotic value of growth index $\gamma_\infty$ of this model is somewhat smaller than that of the $\Lambda$CDM model by approximately $3\%$. This is expected as the agglomerations of DE perturbations usually lowers the value of the growth index \cite{tis}.  It has been also observed that a redshift dependent parametrization of $\gamma$ can provide a more accurate fit for the growth rate of matter clustering in comparison to a constant parametrization. Further, the variation of the growth index may contain crucial information about the underlying gravitational theory. In what follows, we, therefore, analyze the evolution of a time-dependent growth index.

	\begin{figure}[h!]
		\centering
		\subfigure[]{%
			\includegraphics[width=6cm,height=3.5cm]{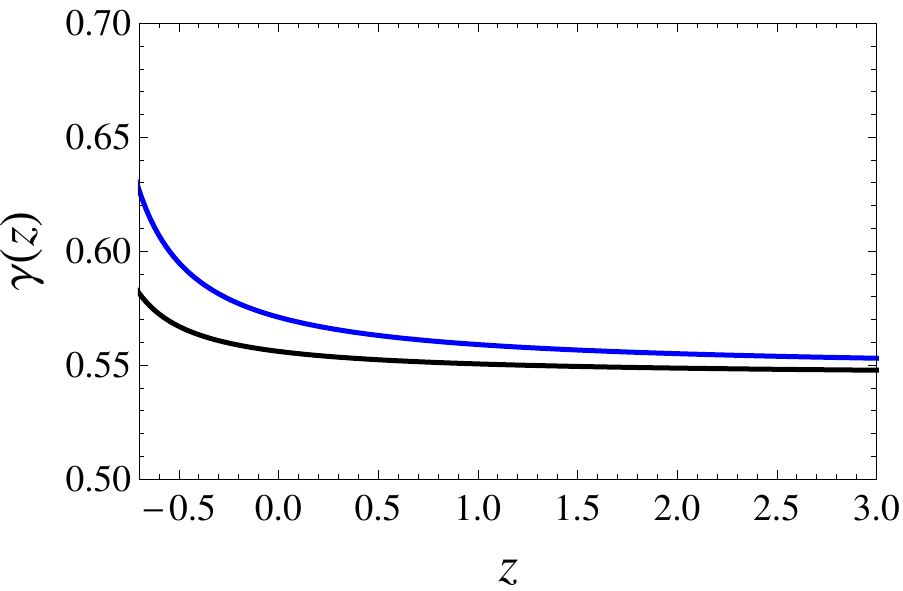}\label{fig_var_gr_ind_ras_lcdm_0_98}}
		\qquad
		\subfigure[]{%
			\includegraphics[width=6cm,height=3.5cm]{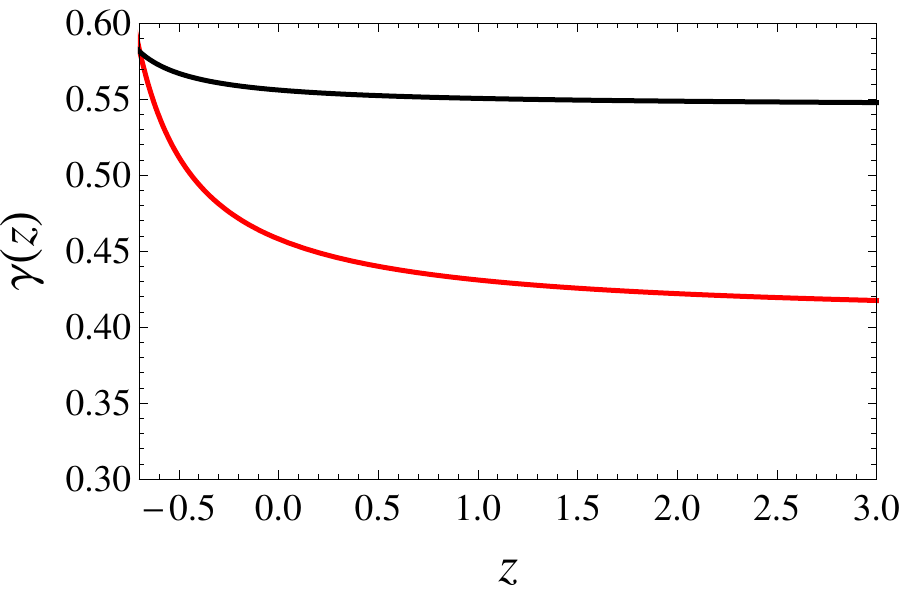}\label{fig_var_gr_ind_ras_lcdm_1_41}}
		\qquad
		\subfigure[]{%
			\includegraphics[width=6cm,height=3.5cm]{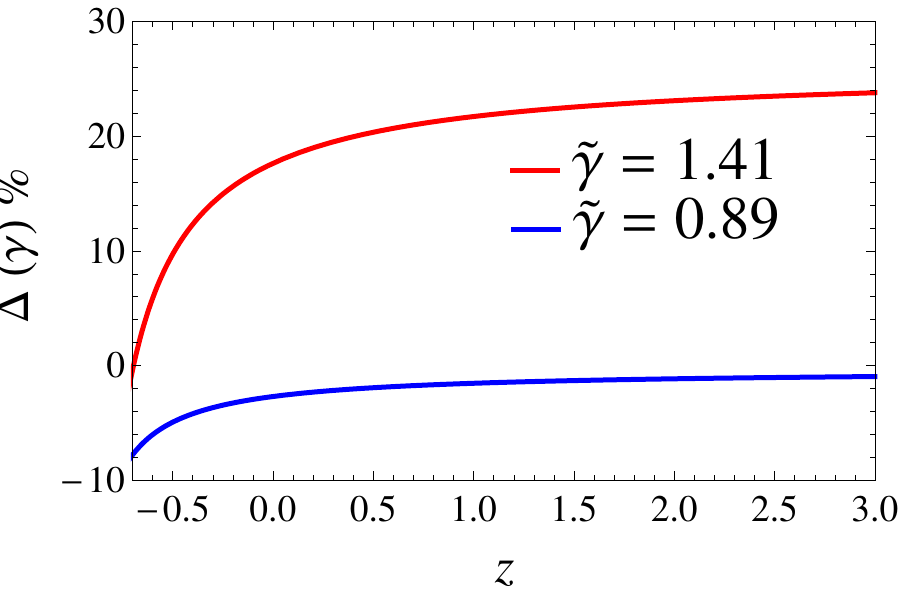}\label{fig_var_gr_ind_ras_lcdm_percent}}
		\caption{The evolution of $\gamma(z)$  as a function of redshift $z$ for Rastall's model  (red and blue) and the $\Lambda$CDM model (black) with $\gtil=0.89$ in panel (a) and $\gtil=1.41$ in  panel (b). In panel (c) we plot the relative difference $\Delta(\gamma_\infty)=[1-\gamma(z)/\gamma^{(\Lambda)}(z)] \%$ of the varying growth index for the Rastall's model with respect to the $\Lambda$CDM model for $\gtil=1.41$ and $0.89$.} 
		\label{fig_var_gr_ind_ras_lcdm}
	\end{figure}

	\subsubsection{Varying growth index}\label{subsec:vary_g_ind}
	Another possibility is the choice of the growth index $\gamma$ which changes with redshift. In this case, we utilize the method of Polarski and Gannouji \cite{pol}. On substituting Eq. \eqref{gr_ind} in Eq. \eqref{df_da} and employing Eq. \eqref{d_Om_d_a} we get
	{\small \begin{equation}\label{dgama_da}
		a \ln (\Omega_m) \frac{d \gamma}{d a}+\Omega^\gamma_m-3\Big(\gamma-\frac{1}{2}\Big) \Big(1+\frac{2}{3} \frac{d \ln E}{d \ln a}\Big)+\frac{1}{2}-\frac{3}{2}\,\mu\, \Omega^{1-\gamma}_m=0\,.
		\end{equation}}
	Evaluating the above equation at the present time which corresponds to a redshift $z=0$ (i.e. $a=1$) we get
	{\small \begin{eqnarray}\label{dgama_da_1}
		-\gamma'(1) \ln \Omega_{m0}+ \Omega^{\gamma(1)}_{0}-3\Big(\gamma(1)-\frac{1}{2}\Big)\,\Big(1+\frac{2}{3}\,\frac{d \ln E}{d \ln a}\Big)_{a=1}\nonumber\\ +\frac{1}{2}-\frac{3}{2}\,\mu_0\,\Omega^{1-\gamma(1)}_{m0}=0\,,~~~~~~~~~~~~~~
		\end{eqnarray}}
	where 
	{\small \begin{eqnarray}
		\mu_0&=&\mu(1)\simeq \gtil+\left(\gtil+3(2-\gtil)\,w_x\right) \nonumber\\&& \Bigg( \frac{\frac{\gtil-1}{2}\frac{\Omega_{m0}}{\Omega_{x0}}+(1+w_x)}{1+\frac{1-\gtil}{2}(1-3w_x)}\Bigg) \Bigg(\frac{\Omega_{x0}}{\Omega_{m0}}\Bigg)\,,\\
		\frac{d \ln E}{d \ln a} \Bigg|_{a=1}&=&- \frac{3 \, \Omega_{m0}\,w_x\,(2\gtil-3)}{3\gtil\,w_x-\gtil-3w_x+3}\,. \label{d_lnE_d_lna1}
		\end{eqnarray}}
	In this work, we consider an approximated solution of Eq. \eqref{dgama_da} expressed as a first order Taylor's expansion about the present epoch i.e. $a(z)=1$ given by
	{\small \begin{equation} \label{gama_tayl}
		\gamma(a)=\gamma_0+\gamma_1 (1-a).
		\end{equation}}
	
	\noindent For large redshift ($z\gg 1$) i.e. $a(z) \simeq 0$, we denote the asymptotic growth index by $\gamma_\infty$, so that $\gamma_\infty \simeq \gamma_0+\gamma_1$ where $\gamma(1)=\gamma_0$. Using Eqs. \eqref{dgama_da_1}, \eqref{d_lnE_d_lna1} and \eqref{gama_tayl}, one can expressed $\gamma_1$ in terms of $\gamma_0$ as
	{\small \begin{eqnarray}
		\gamma_1&=&\frac{1}{\ln \Omega_{m0}}\Bigg[\Omega_{m0}^{\gamma_0}-3\Big(\gamma_0-\frac{1}{2}\Big) \Big(1- \frac{2 \, \Omega_{m0}\,w_x\,(2\gtil-3)}{3\gtil\,w_x-\gtil-3w_x+3} \Big) \nonumber\\&&-\frac{3}{2}\mu_0\Omega^{1-\gamma_0}_{m0}+\frac{1}{2}\Bigg]. \label{gama1_in_gama0}
		\end{eqnarray}}
	
	\noindent Finally, substituting $\gamma_0=\gamma_\infty-\gamma_1$, in Eq. \eqref{gama1_in_gama0} and employing the expression of $\gamma_\infty$ from Eq. \eqref{gam_inf_rast}, we can express $\gamma_0$ and $\gamma_1$ in terms of $\Omega_{m0}$, $\gtil$ and $w_x$.  Note that for various class of models within GR, we have $\vert \gamma_1 (z=0) \vert \lesssim 0.02$. However, this value can be greater in models outside the GR framework \cite{pol1}. Here, if we use parameters values  $\gtil=0.89$, $w_x=-1$ and $\Omega_{m0}=0.296$, then we obtain $(\gamma_0,\gamma_1)=(0.571,-0.024)$. However, if we use parameters values  $\gtil=1.41$, $w_x=-1$ and $\Omega_{m0}=0.296$, then we obtain $(\gamma_0,\gamma_1)=(0.458,-0.054)$.  Also,  for $\Lambda$CDM model where $\gtil=1$, $w_x=-1$ and $\Omega_{m0}=0.3$, we have $(\gamma_0,\gamma_1)=(0.556,-0.011)$. In Figs. \ref{fig_var_gr_ind_ras_lcdm_0_98} and \ref{fig_var_gr_ind_ras_lcdm_1_41}, we present a comparison of a varying growth index of this model  and the $\Lambda$CDM model for $(\gtil,w_x)=(0.89,-1)$ and $(1.41,-1)$ respectively. It can be seen that for $\gtil=0.89$, the evolution of growth index is greater than that of $\Lambda$CDM model for smaller redshift and approaches the $\Lambda$CDM for large redshift. However, for a favourable choice of model parameters i.e. $\gtil=1.41$, $w_x \simeq -1$, we see that the growth index is smaller than that of the $\Lambda$CDM model throughout the evolutionary history of the universe and ultimately approach to that of the $\Lambda$CDM (see Fig. \ref{fig_var_gr_ind_ras_lcdm_1_41}). From Fig. \ref{fig_var_gr_ind_ras_lcdm_percent} it can be also seen that at the present epoch, the growth index of this model is smaller than the $\Lambda$CDM by about $18\%$. Hence, it is interesting to note here that for $w_x=-1$, even though evolution of matter perturbations coincides with that of the $\Lambda$CDM,  however the evolution of the growth index varies from the $\Lambda$CDM for different values of $\gtil$. Therefore, one need to impose tight constraints on the parameter $\gtil$, in order to test possible deviations of this theory from GR.

	\section{Conclusion}\label{sec:conc}
	In the present work, motivated by an open debate on the comparison between Rastall gravity (with the vacuum form of DE) and GR, we analyze the behaviour of the Rastall's cosmological model at the background and linear perturbation level. At the background level, this model shows no deviation from the $\Lambda$CDM model at late time irrespective of the value of Rastall's parameter $\gtil$. However, at the level of linear perturbation, we observed that even though the evolution equation of the matter overdensities for this model coincides with that of the $\Lambda$CDM, the constant value of growth index of matter perturbations shows significant deviation from the $\Lambda$CDM. Furthermore, we obtained that the varying growth index is smaller than the corresponding case of the $\Lambda$CDM for $\gtil>1$. In particular, for best-fit parameters values, we obtained that the asymptotic value of the growth index is about $3\%$ smaller than the $\Lambda$CDM, but the varying growth index is about $18\%$ smaller than the $\Lambda$CDM at the present epoch. Therefore, the growth index value is outside the range of GR based DE models.  Although we do not perform an exhaustive study of the Rastall model, the results of the present work, indeed shed some light on the non-equivalence of Rastall gravity and GR. Further, the smaller value of the growth index for this model is also expected due to the presence of DE perturbations. The influence of DE perturbations on the behaviour of the growth index is crucial, particularly in the light of future generation surveys to test the existence of DE perturbations in the observed Universe \cite{sapone}. Moreover, the agglomeration of DE perturbations may influence the early structure formation at the level of galaxies and cluster of galaxies \cite{batis2}.  The results obtained here also signify the possible deviation of the present model from the $\Lambda$CDM at the nonlinear regime of cosmic perturbations as claimed in Ref. \cite{batis2}.  A deeper analysis along this line is a subject of future work to further test the viability of Rastall theory in comparison to GR with upcoming precise observational data.
	
	\section*{Acknowledgments}
	JD was supported by the Core research grant of  SERB, Department of Science and Technology India (File No.CRG/2018/001035) and the Associate program of IUCAA. The authors thank the referee for constructive suggestions which lead to the improvement of the work.

\end{document}